\documentstyle[prl,multicol,aps]{revtex}

\newcommand{\rstr}{\hbox{$\vert\mkern-4.8mu\hbox{\rm\`{}}\mkern-3mu$}}

\title{On the Additivity of the Entanglement of Formation}
\author{Fabio Benatti{$^1$}, Heide Narnhofer{$^2$}}
\address{$^1$ Dep. Theor. Phys.,
              University. di Trieste,
              Strada Costiera 11, I-34100, Trieste, Italy \\
         $^2$ Inst. Theor. Phys.,
              University of Vienna,
              Boltzmanngasse 5, A-1090, Wien, Austria 
              \hfill}
\begin{document}
\maketitle

\begin{abstract}
We study whether the entanglement of formation is additive over tensor products 
and derive a necessary and sufficient condition for optimality of vector states
that enables us to show additivity in two special cases.
\smallskip

\end{abstract}

\begin{multicols}{2} \narrowtext

Entanglement plays a crucial role
in teleportation and quantum cryptography and is currently the focus of 
investigations in the developing 
field of quantum information~\cite{1}--\cite{6}.
Quantifying entanglement in a satisfactory way is a major issue in quantum
information; there is a list of minimal desiderata  and the available proposals 
have been proved to comply with all of them but for additivity
for which only numerical support exists~\cite{3}.

In this letter we are concerned with the entanglement of
formation, defined by~\cite{1}
\begin{equation}
E_f(\rho;M_1)=\min\Bigl\{\sum_{i\in I}p_i\, 
S\bigl(\sigma_i\rstr_{M_1}
\bigr)\Bigr\}\ ,\label{1}
\end{equation}
where $\rho$ is a generic mixed state on the tensor product of two full matrix
algebras $M_1\otimes M_2$, the minimum is computed over all decompositions 
$\rho=\sum_{i\in I}p_i\,\sigma_i$ of $\rho$ into pure states
$\sigma_i$ on $M_1\otimes M_2$ and
$S\bigl(\sigma_i\rstr_{M_1}\bigr)$ is the von Neumann entropy of 
the restriction of $\sigma_i$ onto $M_1$.

The question we want to address  
is whether 
\begin{equation}
E_f(\rho\otimes\rho;M_1\otimes M_1)=2\,
E_f(\rho;M_1)\ ,
\label{2}
\end{equation}
where $M_1\otimes M_1$ is short for $M_1\otimes1_2\otimes M_1\otimes 1_2$.

Additivity or its failure will have a quantum information theoretic 
counterpart; there is indeed a connection
between~(\ref{1}) and the maximal accessible information $I(\rho)$ of a 
quantum source described by a mixed state 
$\rho$ on a matrix algebra $M$~\cite{6}--\cite{8}.
If $\rho=\sum_{\ell\in L}q_\ell\rho_\ell$, then
$I(\rho):=\sup_B I_B(\rho)$, where
\begin{eqnarray}
I_B(\rho)=&-&\sum_{i\in I}({\rm Tr}(\rho\,b_i))\,\log({\rm
Tr}(\rho\,b_i)\nonumber\\
&+&\sum_{\ell\in L}q_\ell\,
\sum_{i\in I}({\rm Tr}(\rho_\ell\,b_i))\,\log({\rm Tr}(\rho_\ell\,b_i)\ ,
\label{3}
\end{eqnarray}
the maximum being 
computed over all choices 
$B=\{b_i\}$ of positive $b_i\in M$ such that  
$\sum_{i\in I}b_i=1_M$.
Optimal choices correspond to optimal detection of the
classical information carried by the quantum states $\rho_\ell$. 

Since $q_\ell\rho_\ell\leq\rho$ there exists a unique choice of
operators $0<a_\ell\in M$, $\ell\in L$, with
$\sum_{\ell\in L}a_\ell=1$, such that
\begin{equation}
q_\ell\rho_\ell=\sqrt{\rho}a_\ell\sqrt{\rho}\ ,\quad 
q_\ell={\rm Tr}(\rho a_\ell)\ .
\label{4}
\end{equation}
Let $\cal A$ be a commutative $L$ dimensional 
algebra with identity $1_{\cal A}$ and orthogonal projectors $A_\ell$ with 
$\sum_\ell A_\ell=1_{\cal A}$.
The map $\gamma_{\cal A}:{\cal A}\mapsto M$ obtained by linear extension of 
$A_\ell\mapsto\gamma_{\cal A}(A_\ell)=a_\ell$ is 
positive and $\gamma_{\cal A}(1_{\cal A})=1_M$.
Therefore, given any state $\sigma$ on $M$, the linear functional 
$\sigma\circ\gamma_{\cal A}:{\cal A}\mapsto {\bf C}$,
$\sigma\circ\gamma_{\cal A}(A_\ell)={\rm Tr}(\sigma\,a_\ell)$
defines a state on $\cal A$.
Using~(\ref{4}) and the cyclicity of the trace, 
\begin{equation}
{\rm Tr}(\rho_\ell\,b_i)={{\rm Tr}(\rho\,b_i)\over{\rm Tr}(\rho\,a_\ell)}\,
{\rm Tr}(\sigma^B_i\,a_\ell)\ ,\
\sigma^B_i:={\sqrt{\rho}b_i\sqrt{\rho}\over{\rm Tr}(\rho\,b_i)}\ .
\label{5}
\end{equation}
Setting $p^B_i={\rm Tr}(\rho\,b_i)$,~(\ref{3}) becomes
\begin{equation}
I_B(\rho)=S\bigl(\rho\circ\gamma_{\cal A}\bigr)-
\sum_{i\in I}p^B_i S\bigl(\sigma^B_i\circ\gamma_{\cal A}\bigr)\ .
\label{6}
\end{equation}
Therefore, $I(\rho)$ is the maximum of~(\ref{6}) over all possible
decompositions of $\rho$ into pure states:
\begin{equation}
I(\rho)=S\bigl(\rho\circ\gamma_{\cal A}\bigr)-
\min_{\rho=\sum_ip_i\sigma_i}
\sum_ip_iS\bigl(\sigma_i\circ\gamma_{\cal A}\bigr)\ .
\label{7}
\end{equation}

If $N$ is a subalgebra of $M$, substituting the restrictions
$\rho\rstr_N$, $\sigma_i\rstr_N$ for $\rho\circ\gamma_{\cal A}$,
respectively $\sigma_i\circ\gamma_{\cal A}$,
we obtain the so-called {\it entropy of a subalgebra}~\cite{9}
\begin{equation}
H_\rho(N):=S(\rho\rstr_N)-\min_{\rho=\sum_ip_i\sigma_i}
\sum_ip_i\, S\bigl(\sigma_i\rstr_N\bigr)\ .
\label{8}
\end{equation}
The latter quantity is the building block of an extension of the
Kolmogorov-Sinai dynamical entropy (or entropy per unit time) to
the quantum realm.
According to the above
$\displaystyle
E_f(\rho;M_1)=S\bigl(\rho\rstr_{M_1}\bigr)\,-\,
H_\rho(M_1)$.

As the von Neumann entropy is additive over tensor products, if
additivity fails for the entanglement of formation, it also fails
for the entropy of a subalgebra.
Then, from an information-theoretic point of view, 
we would deem possible to extract
more information about the tensor product of two states over two 
independent subalgebras than that obtainable
from the two of them independently~\cite{10,11}.

In the following we try to use some of the properties of $H_\rho(N)$ to
investigate the general question whether
\begin{equation}
E_f(\rho\otimes\sigma;M_1\otimes M_3)=E_f(\rho;M_1)+
E_f(\sigma;M_3)\ ,
\label{9}
\end{equation}
where $\rho$ and $\sigma$ are states on the (finite dimensional)
algebras $M_1\otimes M_2$, respectively $M_3\otimes M_4$.

If $E_f(\rho;M_1)$ and $E_f(\sigma;M_3)$ are achieved at
optimal decompositions $\rho=\sum_\ell p_\ell\rho_\ell$
and $\sigma=\sum_j q_j\sigma_j$, the factorized decomposition 
$
\rho\otimes\sigma=\sum_{j,\ell}q_jp_\ell \rho_\ell\otimes\sigma_j
$
contribute to $E_f(\rho\otimes\sigma;M_1\otimes M_3)$ with 
$E_f(\rho_1;M_1)+E_f(\sigma;M_3)$.
However, the latter need not be optimal and the strict inequality
$E_f(\rho\otimes\sigma;M_1\otimes M_3)<
E_f(\rho_1;M_1)+E_f(\sigma;M_3)$ is not excluded.
In fact, a 
decomposition 
\begin{equation}
\rho\otimes\sigma=\sum_i\alpha_i|\psi_i\rangle\langle\psi_i|\ ,\
\alpha_i>0\ ,\ \sum_i\alpha_i=1\ ,
\label{10}
\end{equation}
might be optimal with the $\psi_i$ entangled states over $M_1\otimes M_3$.
Let us consider the Schmidt decomposition
\begin{equation}
|\psi_i\rangle
=\sum_j\beta_{ij}|\phi^{12}_{ij}\rangle\otimes|\phi^{34}_{ij}\rangle\ ,\
\|\psi_i\|=1\ ,\ \beta_{ij}>0\ ,
\label{11}
\end{equation}
where, for fixed $i$ the $|\phi^{12}_{ij}\rangle$'s and 
$|\phi^{34}_{ij}\rangle$'s form orthonormal bases over 
$M_1\otimes M_2$, respectively $M_3\otimes M_4$.
If it held that
\begin{eqnarray}
S\bigl(|\psi_i\rangle\langle\psi_i|\rstr_{M_1\otimes M_3}\bigr)&\geq&
\sum_j\beta_{ij}^2\Bigl(
S\bigl(|\phi^{12}_{ij}\rangle\langle\phi^{12}_{ij}|\rstr_{M_1}\bigr)
\nonumber\\
&+&
S\bigl(|\phi^{34}_{ij}\rangle\langle\phi^{34}_{ij}|\rstr_{M_3}\bigr)\Bigr)
\label{12}
\end{eqnarray}
additivity would follow because tensor-product states 
would then never be worse than correlated ones.

Proving the sufficient condition~(\ref{12}) has so far escaped us; there are
however particular cases
where one can show additivity by using two results obtained 
for the entropy of a subalgebra~(\ref{8}).
Both results concern general properties
of optimal decompositions for $H_\rho(N)$ that we
adapt to the entanglement of formation~\cite{12,13}.
\medskip

{\bf Proposition 1.}\quad
If $\rho$ is a state on $M_1\otimes M_2$, 
$E_f(\rho;M_1)$ is achieved at  
$\rho=\sum_\ell p_\ell\rho_\ell$ and $U$ is a unitary operator on 
$M_1\otimes M_2$, then
$E_f(U^\dagger\rho U;M_1)$ 
is achieved at the optimal decomposition 
$U^\dagger\rho U=\sum_\ell p_\ell U^\dagger\rho_\ell U$.
\smallskip

{\bf Proposition 2.}\quad
Let $\rho$ be a state on  $M_1\otimes M_2$ and
$E_f(\rho;M_1)$ be achieved at $\rho=\sum_\ell p_\ell\rho_\ell$,
That is, $E_f(\rho;M_1)
=\sum_\ell p_\ell\,S\bigl(\rho_\ell\rstr_{M_1}\bigr)$, then
\begin{equation}
E_f(\sigma;M_1)=\sum_jq_jS\bigl(\rho_j\rstr_{M_1}\bigr)\ ,
\label{13}
\end{equation}
where $\sigma=\sum_jq_j\rho_j$ 
is any linear convex combination of optimal states
of $\rho$.
\smallskip

To the above, we add a new property.
With some abuse of notation, we denote by $E_f(\rho;N)$ the 
minimum in~(\ref{8}), even if there is no tensor product structure in $N$.

{\bf Proposition 3.}\quad
Let $|\psi_i\rangle\langle\psi_i|$, $i=1,2$, contribute
to $E_f(\rho;N)$ and denote
\begin{eqnarray*}
&&
\sigma_i:=|\psi_i\rangle\langle\psi_i|\rstr_N\ ,\ i=1,2\ ;\
\sigma_{12}:=|\psi_1\rangle\langle\psi_2|\rstr_N\\ 
&&
\sigma_{ov}(\gamma):=\gamma\sigma_{21}+\gamma^*\sigma_{12}\\
&&
\sigma(\gamma):=|\gamma|^2\sigma_1+\sigma_2-\sigma_{ov}(\gamma)\ ,\
\hat{\sigma}(\gamma):={\sigma(\gamma)\over{\rm Tr}(\sigma(\gamma))}\ .
\end{eqnarray*}
Then, for all complex $\gamma$,
\begin{equation}
{|\gamma|^2S(\sigma_1)+S(\sigma_2)+
{\rm Tr}\Bigl(\sigma_{ov}(\gamma)\log\sigma_1\Bigr)\over
{\rm Tr}(\sigma(\gamma))}
\leq 
S(\hat{\sigma}(\gamma))\ .
\label{14}
\end{equation}

\noindent
Vice versa, if inequality~(\ref{14}) 
holds for all complex $\gamma$, then for all 
$\rho_\lambda=\lambda |\psi_1\rangle\langle\psi_1|+(1-\lambda)
|\psi_2\rangle\langle\psi_2|$, $1\geq \lambda\geq0$, one gets
$E_f(\rho_\lambda;N)=\lambda S(\sigma_1)+(1-\lambda) S(\sigma_2)$.
\smallskip

\noindent
{\bf Proof of Necessity:}\quad
Let $\varepsilon>0$ and set
$$
\rho_{\varepsilon,\gamma}=
(1+\varepsilon|\gamma|^2)|\psi_1\rangle\langle\psi_1|+
\varepsilon(1+\varepsilon|\gamma|^2)|\psi_2\rangle\langle\psi_2|
$$
be a not normalized state on $M$.
As $\psi_i$, $i=1,2$ are optimal, Proposition 2 yields
$$
E_f(\rho_{\varepsilon,\gamma};N)= (1+\varepsilon|\gamma|^2)\, S(\sigma_1)
+ \varepsilon(1+\varepsilon|\gamma|^2)\, S(\sigma_2)\ .
$$
Indeed, in taking the minimum in~(\ref{8}) normalization is not necessary.
With $|\phi_1\rangle:=|\psi_1\rangle+\varepsilon\gamma|\psi_2\rangle$ and
$|\phi_2\rangle:=|\psi_1\rangle-(\gamma*)^{-1}|\psi_2\rangle$, we construct
a new decomposition
$\rho_{\varepsilon,\gamma}=|\phi_1\rangle\langle\phi_1|
+\varepsilon|\gamma|^2|\phi_2\rangle\langle\phi_2|$.

The latter cannot contribute more 
than $E_f(\rho_{\varepsilon,\gamma};N)$;
therefore, 
$E_f(\rho_{\varepsilon,\gamma};N)\leq f(\varepsilon)$,
where
$$
f(\varepsilon):=\|\phi_1\|^2
S\Bigl(|\phi_1\rangle\langle\phi_1|\rstr_N\Bigr)+
\varepsilon|\gamma|^2\|\phi_2\|^2
S\Bigl(|\phi_2\rangle\langle\phi_2|\rstr_N\Bigr)\ . 
$$
Inequality~(\ref{14}) must then hold at 
first order in $\varepsilon$.
\smallskip

\noindent
{\bf Proof of Sufficiency:}\quad
By assumption, inequality~(\ref{14}) holds for all $\gamma$'s.
Thus, choosing $\alpha_i\geq0$ and $\gamma_i$ such that
$\sum_i\alpha_i|\gamma_i|^2=\lambda$, $\sum_i\alpha_i=1-\lambda$ and
$\sum_i\alpha_i\gamma_i=0$, we get
$$
\lambda\,S(\sigma_1)+(1-\lambda)\,S(\sigma_2)\leq
\sum_i\alpha_i\Bigl({\rm Tr}\sigma(\gamma_i)\Bigr)\,
S(\hat{\sigma}(\gamma_i))\ .
$$
In the above, the left hand side is the contribution to
$E_f(\rho_\lambda;N)$ of 
$
\rho_\lambda=\lambda|\psi_1\rangle\langle\psi_1|
+(1-\lambda)|\psi_2\rangle\langle\psi_2|\ ,
$ 
whereas the right hand side is the
contribution of 
\begin{equation}
\rho_\lambda=
\sum_i\alpha_i|\psi_1+\gamma_i\psi_2\rangle\langle\psi_1+\gamma_i\psi_2|\ .
\label{15}
\end{equation}
The latter are the most general decompositions of $\rho_\lambda$; in fact,
$\rho_\lambda$ as an operator acts on the two-dimensional subspace spanned by
the linearly independent vectors $\psi_i$, $i=1,2$.
Hence, the result follows.
\smallskip

With the help of Propositions 1, 2 and 3 we can now prove additivity in some 
special cases.
\smallskip

{\bf Case 1:}\quad
In~(\ref{11}) the state $\sigma$ factorizes over $M_3\otimes M_4$:
$\sigma=\sigma_3\otimes\sigma_4$.

Let $E_f(\rho\otimes\sigma;M_1\otimes M_3)$ be achieved at an optimal 
decomposition
made of states $|\psi_i\rangle$ entangled over $M_1\otimes M_3$.
Let us consider the Schmidt decompositions 
$|\psi_i\rangle=\sum_jc_{ij}|\phi^{13}_{ij}\rangle
\otimes|\phi^{24}_{ij}\rangle$ with
$c_{ij}\geq0$ and
$|\phi^{13}_{ij}\rangle$ and $|\phi^{24}_{ij}\rangle$ 
forming, for each fixed $i$,  
orthonormal bases in the first and third factor, respectively. 
Thus,
\begin{eqnarray}
\rho\otimes\sigma_3\otimes\sigma_4&=&
\sum_i\alpha_i|\psi_i\rangle\langle\psi_i|
\nonumber\\
&=&\sum_{ijk}\alpha_ic_{ij}c_{ik}|\phi^{13}_{ij}\rangle\langle\phi^{13}_{ik}|
\otimes|\phi^{24}_{ij}\rangle\langle\phi^{24}_{ik}|\ .
\label{16}
\end{eqnarray}
Further, let $|\chi^3_\ell\rangle$ be eigenvectors of $\sigma_3$ and consider
the unitary operator ($\in M_3$)
$$
\hat{U}_\ell=1-(1-i)P_\ell\ ,\quad 
P_\ell:=|\chi^3_\ell\rangle\langle\chi^3_\ell|\ .
$$
From Proposition 1 it follows that the vectors
$U_\ell|\psi_i\rangle$, 
$U_\ell=1_1\otimes 1_2\otimes \hat{U}_\ell\otimes 1_4$, also give 
$E_f\bigl(\rho\otimes \sigma;M_1\otimes M_3\bigr)$.
Let us concentrate on $|\psi_1\rangle$; together with
$U_\ell|\psi_1\rangle$, they have to satisfy~(\ref{14}) 
for all $\gamma$.
Then, according to the notation of Proposition 3,
\begin{eqnarray}
\sigma_1&=&\sum_jc_{1j}^2|\phi^{13}_{1j}\rangle\langle\phi^{13}_{1j}|
\rstr_{M_1\otimes M_3}
\label{17}\\
\sigma_2&=&U_\ell\sigma_1U_\ell^\dagger\ ,\quad
\sigma_{ov}(\gamma)=
\gamma U_\ell\sigma_1+\gamma^*\sigma_1U_\ell^\dagger\ .
\label{18}
\end{eqnarray}

Taking $\gamma=1$, it follows that
$\displaystyle
\hat{\sigma}(1)={P_\ell\sigma_1 P_\ell\over{\rm Tr}(P_\ell\sigma_1)}$ and
$\sigma_{ov}(1)=2\sigma_1-(1-i)P_\ell\sigma_1-(1+i)\sigma_1 P_\ell$.
Inequality~(\ref{14}) thus becomes
\begin{equation}
-2{\rm Tr}\Bigl(P_\ell\sigma_1\log\sigma_1\Bigr)\leq{\rm
Tr}\Bigr(P_\ell\sigma_1\Bigr)\,S\Bigl(\hat{\sigma}(1)\Bigr)\ .
\label{19}
\end{equation}
We  develop
$|\phi^{13}_{1j}\rangle
=\sum_p\beta^j_{p\ell}|\chi^1_p\rangle\otimes|\chi^3_\ell\rangle$, 
along an orthonormal basis for the factor $M_1$, then, by means
of the spectral decomposition~(\ref{17}), 
setting
$\Delta_{j\ell}:=\langle\phi^{13}_j|P_\ell|\phi^{13}_j\rangle=
\sum_p|\beta^j_{p\ell}|^2$, we get
$\displaystyle
P_\ell\sigma_1 P_\ell=\Bigl(\sum_jc_{1j}^2\Delta_{j\ell} Q_{j\ell}\Bigr)\otimes
P_\ell
$, where
$
Q_{j\ell}:=|\hat{\chi}^1_{j\ell}\rangle\langle\hat{\chi}^1_{j\ell}|
$
and 
$\displaystyle
|\hat{\chi}^1_{j\ell}\rangle=
\sum_p{\beta^j_{p\ell}\over\Delta_{j\ell}}|\chi^1_p\rangle
$.

Insertion in~(\ref{19}) leads to 
\begin{eqnarray*}
0&\geq&\sum_jc_{1j}^2\Delta_{j\ell}\log{\Delta_{j\ell}\over c_{1j}^2
{\rm Tr}(P_\ell\sigma)}
\\
&\geq&
\sum_jc_{1j}^2\Bigl(\Delta_{j\ell}-c_{1j}^2{\rm Tr}(P_\ell\sigma)\Bigr)\ ,
\end{eqnarray*}
the latter inequality coming from $x\log x/y\geq x-y$ and holding 
for all orthogonal projectors $P_\ell$.
Since $\sum_jc_{1j}^2=1$ and $\sum_\ell\Delta_{j\ell}=1$,
summing over $\ell$ we get that $c_{1j}=1$ for one $j$ and $c_{1k}=0$ if $k\neq j$.
Thus, the supposed optimal vectors $|\psi_i\rangle$ must be of the form
$|\psi_i\rangle=|\phi^{13}_i\rangle\otimes|\phi^{24}_i\rangle$ and the 
supposed optimal decomposition~(\ref{16}) must reduce to
\begin{equation}
\rho\otimes\sigma_3\otimes\sigma_4=\sum_i\alpha_i
|\phi^{13}_i\rangle\langle\phi^{13}_i|\otimes
|\phi^{24}_i\rangle\langle\phi^{24}_i|\ .
\label{20}
\end{equation}

Tracing over $M_2\otimes M_4$ with respect to the Schmidt decompositions
$|\phi^{13}_i\rangle=
\sum_j\delta_{ij}^{13}|\phi^1_{ij}\rangle\otimes|\phi^3_{ij}\rangle$
and 
$|\phi^{24}_i\rangle=
\sum_j\delta_{i\ell}^{24}|\phi^2_{i\ell}\rangle\otimes|\phi^4_{i\ell}\rangle
$, orthogonality yields
\begin{equation}
\rho=\sum_i\alpha_i\sum_{j\ell}(\delta^{13}_{ij})^2(\delta^{24}_{i\ell})^2\,
|\phi^1_{ij}\rangle\langle\phi^{1}_{ij}|\otimes
|\phi^2_{i\ell}\rangle\langle\phi^2_{i\ell}|\ .
\label{21}
\end{equation}
We thus conclude that a decomposition of $\rho\otimes\sigma_3\otimes\sigma_4$
as in~(\ref{16}) can be optimal with
respect to $M_1\otimes M_3$ only if $\rho$ is not entangled over 
$M_1\otimes M_2$,
in which case $E_f(\rho\otimes\sigma_3\otimes\sigma_4;M_1\otimes M_3)=0$
is obviously additive.
If $\rho$ is entangled over $M_1\otimes M_2$, the contradiction is avoided only
if the optimal decompositions have the form 
\begin{equation}
\rho\otimes\sigma_3\otimes\sigma_4=\sum_i\alpha_i
|\phi^{12}_i\rangle\langle\phi^{12}_i|\otimes
|\phi^{34}_i\rangle\langle\phi^{34}_i|\ .
\label{22}
\end{equation}
Thus, the optimal states cannot carry any entanglement over 
$M_1\otimes M_3$ and additivity follows.
\smallskip

The second case we want to discuss is somewhat the opposite of the 
previous one where we proved that optimal projections for the tensor products 
are products of optimal projectors for the factors.
In the second case, we want to show that putting together couples of
optimal projectors for
the factors we get optimal decompositions.

{\bf Case 2:}\quad
we consider the state 
\begin{equation}
\rho_\lambda=\lambda\rho+(1-\lambda)\hat{\rho}
\label{23}
\end{equation}
on
$M_1\otimes M_2\otimes M_3\otimes M_4$, where 
$\rho:=|\phi^{12}\rangle\langle\phi^{12}|
\otimes|\phi^{34}\rangle\langle\phi^{34}|
$ and 
$\hat{\rho}:=|\hat{\phi}^{12}\rangle\langle\hat{\phi}^{12}|\otimes
|\hat{\phi}^{34}\rangle\langle\hat{\phi}^{34}|
$.

Let
$|\phi^{12}\rangle$ and $|\hat{\phi}^{12}\rangle$ be optimal vectors
for some state $\rho$ on $M_1\otimes M_2$ relative to $M_1$ and
$|\phi^{34}\rangle=|\phi^3\rangle\otimes|\phi^4\rangle$,
$|\hat{\phi}^{34}\rangle=|\hat{\phi}^3\rangle\otimes|\hat{\phi}^4\rangle$
on $M_3\otimes M_4$ so that 
$E\Bigl(\rho_\lambda;M_3\Bigr)=0$.
The contribution to 
$E\Bigl(\rho_\lambda;M_1\otimes M_3\Bigr)$ of the decomposition~(\ref{23}) 
is thus
\begin{equation}
E_\lambda:=\lambda S\Bigl(|\phi^{12}\rangle\langle\phi^{12}|\rstr_{M_1}\Bigr)
+(1-\lambda)
S\Bigl(|\hat{\phi}^{12}\rangle\langle\hat{\phi}^{12}|\rstr_{M_1}\Bigr)
\label{24}
\end{equation}
and we want to prove that this is the best we can have.

We proceed as follows: as for~(\ref{15}),
a general decomposition of $\rho_\lambda$ is of the form
$\rho_\lambda=\sum_i\alpha_i|\psi_i\rangle\langle\psi_i|$
where
$|\psi_i\rangle=|\phi^{12}\otimes\phi^{34}\rangle+
\gamma_i|\hat{\phi}^{12}\otimes\hat{\phi}^{34}\rangle$,
with $\alpha_i>0$ and  
\begin{equation}
\sum_i\alpha_i=\lambda\ ,\
\sum_i\alpha_i|\gamma_i|^2=1-\lambda\ ,\
\sum_i\alpha_i\gamma_i=0\ .
\label{25}
\end{equation}
We now set $b:=\langle\phi^4|\hat{\phi}^4\rangle$, $a:=\sqrt{1-|b|^2}$ and
construct the normalized vector state 
$\displaystyle
|\psi^4\rangle:={|\hat{\phi}^4\rangle-b|\phi^4\rangle\over a}$ such that
$\langle\psi^4|\phi^4\rangle=0$.
We can thus rewrite
\begin{equation}
|\psi_i\rangle=a_i|\phi_i^{123}\otimes\phi^4\rangle+
a\gamma_i|\hat{\phi}^{12}\otimes\hat{\phi}^3\otimes\psi^4\rangle\ ,
\label{26}
\end{equation}
where
\begin{eqnarray*}
&&
|\phi_i^{123}\rangle:={|\phi^{12}\otimes\phi^3\rangle
+b\gamma_i|\hat{\phi}^{12}\otimes\hat{\phi}^3\otimes\phi^4\rangle\over a_i}
\nonumber\\
&&
a^2_i:=1+|b|^2|\gamma_i|^2+2{\cal R}e\bigl(b\gamma_i
\langle\hat{\phi}^{12}|\phi^{12}\rangle\langle\hat{\phi}^3|\phi^3\rangle\bigr)
\ .
\end{eqnarray*}
With  
$\displaystyle
|\hat{\psi}_i\rangle:={|\psi_i\rangle\over\sqrt{\delta_i}}$, 
$\delta_i:=a_i^2+a^2|\gamma_i|^2$,
the decomposition~(\ref{23}) reads
$\rho_\lambda=\sum_i\alpha_i\delta_i|\hat{\psi}_i\rangle\langle\hat{\psi}_i|$. 

The contribution of the latter to the entanglement of formation 
$E_f(\rho_\lambda;M_1\otimes M_3)$ is
\begin{equation}
E:=\sum_i\alpha_i\delta_i\,
S\Bigl(|\hat{\psi}_i\rangle\langle\hat{\psi}_i|\rstr_{M_1\otimes
M_3}\Bigr)\ .
\label{27}
\end{equation}
From the orthogonality of $\psi^4$ and $\phi^4$ it follows that
$$
|\hat{\psi}_i\rangle\langle\hat{\psi}_i|\rstr_{M_1\otimes M_3}=
{a_i^2\over\delta_i}\sigma^{123}_i+{a^2|\gamma_i|^2\over\delta_i}
\sigma^{123}\ ,
$$
where
\begin{eqnarray}
\sigma^{123}_i&:=&
|\phi^{123}_i\rangle\langle\phi^{123}_i|\rstr_{M_1\otimes M_3}
\label{si}
\\
\sigma^{123}&:=&|\hat{\phi}^{12}\rangle\langle\hat{\phi}^{12}|\rstr_{M_1}
\otimes|\hat{\phi}^3\rangle\langle\hat{\phi}^3|\rstr_{M_3}\ .\nonumber
\end{eqnarray}
Concavity of the von Neumann entropy yields
\begin{eqnarray}
&&E\geq\sum_i\alpha_i\Bigl\{
a_i^2\,S\Bigl(\sigma_i^{123}\rstr_{M_1\otimes M_3}\Bigr)
\nonumber\\
&&\hskip 3cm
a^2\,|\gamma_i|^2\,
S\Bigl(|\hat{\phi}^{12}\rangle\langle\hat{\phi}^{12}|\rstr_{M_1}\Bigr)
\Bigr\}\ .
\label{28}
\end{eqnarray}
As done before, we construct the normalized vector 
$\displaystyle
|\psi^3\rangle:={|\hat{\phi}^3\rangle-d|\phi^3\rangle\over c}$,
such that $\langle\psi^3|\phi^3\rangle=0$ where 
$d:=\langle\phi^3|\hat{\phi}^3\rangle$, $c:=\sqrt{1-|d|^2}$,
and
\begin{eqnarray}
|\phi^{123}_i\rangle&:=&{b_i|\psi^{12}_i\otimes\phi^3\rangle+
bc\gamma_i|\hat{\phi}^{12}\otimes\psi^3\rangle\over a_i}
\nonumber\\
|\psi^{12}_i\rangle&:=&{|\phi^{12}\rangle
+bd\gamma_i|\hat{\phi}^{12}\rangle\over b_i}
\label{29}\\
b^2_i&:=&1+|b|^2|d|^2|\gamma_i|^2+2{\cal R}e\bigl(
bd\gamma_i\langle\hat{\phi}^{12}|\phi^{12}\rangle\bigr)\ .
\nonumber
\end{eqnarray}
Introducing the Schmidt decompositions over $M_1\otimes M_2$:
$|\psi^{12}_i\rangle=\sum_jc_{ij}|\phi^1_{ij}\rangle\otimes|\phi^2_{ij}\rangle$,
$|\hat{\phi}^{12}\rangle
=\sum_\ell d_\ell|\hat{\phi}^1_\ell\rangle\otimes|\hat{\phi}^2_\ell\rangle$,
and setting
$\rho_1:=|\psi_i^{12}\rangle\langle\psi_i^{12}|\rstr_{M_1}$,
$\rho_2:=|\hat{\phi}^{12}\rangle\langle\hat{\phi}^{12}|\rstr_{M_1}$,
because of the orthogonality of $\phi^3$ and $\psi^3$, the state
$\sigma^{123}_i$ in~(\ref{si}) restricted to $M_1\otimes M_3$
can be represented as 
\begin{eqnarray*}
\sigma^{123}_i&=&{1\over a_i^2}
\pmatrix{
b_i^2\rho_1&
cb_ib^*\gamma_i^*\sqrt{\rho_1}\,V\sqrt{\rho_2}\cr
cb_ib\gamma_i\sqrt{\rho_2}\,V^\dagger\sqrt{\rho_1}&
|b|^2c^2|\gamma_i|^2\rho_2}
\nonumber\\ 
&=&{1\over a_i^2}
\pmatrix{
b_i\sqrt{\rho_1}\,V&0\cr
cb\gamma_i\sqrt{\rho_2}&0}\,
\pmatrix{b_iV^\dagger\sqrt{\rho_1}&cb^*\gamma_i^*\sqrt{\rho_2}\cr
0&0}\ ,
\end{eqnarray*}
where $V:=\sum_{j,\ell}\langle\hat{\phi}^2_\ell\,
|\phi^{12}_{ij}\rangle|\phi^1_{ij}\rangle
\langle\hat{\phi}^1_\ell|$ is a unitary operator
and $\sqrt{\rho_1}\,V\sqrt{\rho_2}=
|\psi^{12}_i\rangle\langle\hat{\phi}^{12}|\rstr_{M_1}$.

Since $\sigma^{123}_i=A^\dagger A$ has the same entropy as
$\displaystyle
AA^\dagger=
{1\over a_i^2}
\pmatrix{
b_i^2V^\dagger\rho_1V+|b|^2c^2|\gamma_i|^2\rho_2&0\cr&\cr0&0}
$,
concavity and invariance under unitary transformations of the von
Neumann entropy yield 
$$
S\Bigl(\sigma^{123}_i\Bigr)=S\Bigl(AA^\dagger\Bigr)\geq
{b_i^2\over a_i^2}S\Bigl(\rho_1\Bigl)+{c^2|b|^2|\gamma_i|^2
\over a_i^2}S\Bigl(\rho_2\Bigr)\ ,
$$
whence~(\ref{28}) becomes
\begin{equation}
E\geq\sum_i\alpha_i\Bigl\{b_i^2S(\rho_1)
+|\gamma_i|^2(a^2+c^2|b|^2)S(\rho_2)\Bigl\}\ .
\label{30}
\end{equation}

Since we assumed the states $|\phi^{12}\rangle$ and $|\hat{\phi}^{12}\rangle$
in~(\ref{29}) to be optimal for some state on $M_1\otimes M_2$ when
restricted to $M_1$, we can use the necessary condition~(\ref{14}).
According to the notation of Proposition 3, we have
$\sigma_1=\rho_2$, $\sigma_2=|\phi^{12}\rangle\langle\phi^{12}|\rstr_{M_1}$,
$\gamma=\gamma^*_ib^*d^*$, $\hat{\sigma}(\gamma)=\rho_1$ and
$$
\sigma_{ov}(\gamma)=-b^*d^*\gamma_i^*\sqrt{\rho_2}\,V^\dagger\sqrt{\rho_1}-
bd\gamma_i\sqrt{\rho_1}\,V\sqrt{\rho_2}\ .
$$
From~(\ref{14}) and the conditions~(\ref{25}) 
it follows that
\begin{eqnarray}
E&\geq&\sum_i\alpha_i\Bigl\{\,
S\Bigl(|\phi^{12}\rangle\langle\phi^{12}|\rstr_{M_1}\Bigr)
+|\gamma_i|^2\,S\Bigl(|\hat{\phi}^{12}\rangle
\langle\hat{\phi}^{12}|\rstr_{M_1}\Bigr)\nonumber\\
&&\hskip .5cm
-{\rm Tr}\Bigl(\sigma_{ov}(b^*d^*\gamma_i^*)\log\rho_2\Bigr)\Bigr\}=E_\lambda
\ ,
\label{31}
\end{eqnarray}
where $E_\lambda$ is the contribution~(\ref{24}) to the entanglement 
of formation $E\Bigl(\rho_\lambda;M_1\otimes M_3\Bigr)$ of the 
decomposition~(\ref{23}), which turns out then to be already optimal.
\smallskip

In this letter we have derived a necessary and sufficient condition for
the optimality of two vector states and showed its usefulness
by proving additivity in two cases.
While in case 1. additivity  was rather expected because of the tensor-product 
state $\rho\otimes\sigma$, it was less so in case 2. 
We requested, however, additional properties on the state
structure over $M_3\otimes M_4$: 
in case 1. factorization of $\sigma=\sigma_3\otimes\sigma_4$ and 
in case 2. factorization into pure states of the optimal decomposers.
In both cases the state was thus separable with respect to $M_3\otimes M_4$.

\end{multicols}

\begin{thebibliography}{99}
\bibitem{1}
C.H. Bennet, D.P. DiVincenzo, J.A. Smolin, and W.K. Wooters, Phys. Rev. A 
{\bf 54}, 3824 (1996)
\bibitem{2}
S. Hill, and W.K. Wootters, Phys. Rev. Lett. {\bf 78}, 5022 (1997)
\bibitem{3}
W.K. Wootters, Phys. Rev. Lett. {\bf 80}, 2245 (1998)
\bibitem{4}
V. Vedral, and M.B. Plenio, Pys. Rev. A{\bf 57}, 1619 (1998)
\bibitem{5}
M. Horodecki, P. Horodecki, R. Horodecki, lanl e-print quant-ph/9908065
\bibitem{6}
G.G. Amosov, A.S. Holevo, and R.F. Werner, lanl e-print math-ph/0003002
\bibitem{7}
A. S. Holevo, Prob. Inf. Transmission USSR {\bf 9} 31 (1993)
\bibitem{8} 
F. Benatti, J. Math. Phys. {\bf 37} 5244 (1996) 
\bibitem{9}
A. Connes, H. Narnhofer, and W. Thirring, Comm. Math. Phys. {\bf 112}, 
691 (1987);
H. Narnhofer, and W. Thirring: Fizika {\bf 17}, 257 (1985) 
\bibitem{10}
A. Peres, and W.K. Wootters, Phys. Rev. Lett. {\bf 66}, 1119 (1991)
\bibitem{11}
N. Gisin and S. Popescu, lanl e-print quant-ph/9901072 
\bibitem{12}
F. Benatti, H. Narnhofer, and A. Uhlmann, Rep. Math. Phys. {\bf 38}, 123 (1996)
\bibitem{13}
F. Benatti, H. Narnhofer, and A. Uhlmann, Lett. Math. Phys. {\bf 47}, 237 (1999)



\end{thebibliography}
\end{document}